# DISTANCE-BASED CLUSTERING OF SPARSELY OBSERVED STOCHASTIC PROCESSES, WITH APPLICATIONS TO ONLINE AUCTIONS

By Jie Peng and Hans-Georg Müller[1]

*University of California, Davis*

We propose a distance between two realizations of a random process where for each realization only sparse and irregularly spaced measurements with additional measurement errors are available. Such data occur commonly in longitudinal studies and online trading data. A distance measure then makes it possible to apply distance-based analysis such as classification, clustering and multidimensional scaling for irregularly sampled longitudinal data. Once a suitable distance measure for sparsely sampled longitudinal trajectories has been found, we apply distance-based clustering methods to eBay online auction data. We identify six distinct clusters of bidding patterns. Each of these bidding patterns is found to be associated with a specific chance to obtain the auctioned item at a reasonable price.

**1. Introduction.** The goal of cluster analysis is to group a collection of subjects into clusters, such that those falling into the same cluster are more similar to each other than those in different clusters. Therefore, a measure of similarity or dissimilarity between subjects is a necessary ingredient for clustering. A metric defined on the subject space is one way to obtain dissimilarities, simply using the distance between two subjects as a measure of dissimilarity. While one can readily choose from a variety of well-known metrics for the case of classical multivariate data, or for functional data that are in the form of continuously observed trajectories, finding a suitable distance measure for irregularly observed data can be a challenge. One such situation which we study here occurs in the commonly encountered case of irregularly and sparsely observed longitudinal data, with online auction data a prominent example [Shmueli and Jank (2005), Jank and Shmueli (2006), Shmueli, Russo and Jank (2007), Liu and Müller

Received October 2007; revised April 2008.
[1]Supported in part by NSF Grants DMS-03-54448 and DMS-05-05537.
*Key words and phrases.* Bidder trajectory, clustering of trajectories, functional data analysis, metric in function space, multidimensional scaling.

(2008)]. As an example, a snapshot of an eBay auction history for a Palm Personal Digital Assistant is shown in Figure 1. In this paper the focus is on a traditional clustering framework, where it is assumed that each subject belongs to exactly one cluster. There are alternative clustering ideas such as soft clustering [Eroscheva and Fienberg (2005)] or mixed membership clustering [Eroscheva, Fienberg and Lafferty (2004)]. For example, in Eroscheva, Fienberg and Joutard (2007), functional disability data are analyzed by a grade of membership model, which allows subjects to have partial membership in several mixture components at the same time.

Online auctions are generating increasingly large amounts of data for which analysis tools are still scarce and eBay is one of the biggest online auction marketplaces. The eBay auction shown in Figure 1 is an example of the type of single-item auctions on which we focus. These are 7-day auctions set up as second-price auctions. eBay uses a proxy bidding system where bidders submit the maximum amounts that they are willing to pay for the item being auctioned, referred to as *WTP—willingness to pay amounts*, and the proxy system automatically increases each bidder's bid by a minimum increment (which is relative to the current highest bid and set by eBay), until either the bidder's maximum has been reached, or the bidder has the current highest bid. During an auction, a bidder can submit as many WTP amounts as desired. The winning bidder is determined according to who has submitted the highest bid at the end of the auction. The price the winning bidder pays corresponds to the second highest bid, plus an increment [Shmueli and Jank (2005)]. We refer to the series of all WTP bids, including the times within the auction at which these were submitted, as the "bid history" of a particular auction. It is noteworthy that consecutive WTP amounts can decrease and therefore are not constrained to be monotone increasing, since all but the highest current WTP amounts are visible during an ongoing auction.

For studying bidding behaviors in eBay auctions, we will focus here on the WTP amounts, as these reflect the intentions of bidders and therefore capture bidder behavior. A characteristic of online auction data is that the times when bids are placed are sparsely located within the time domain of an auction (7 days in our examples), as many bidders submit only very few bids (one or two) during a given auction, and the timing of their bids is irregular. It is well known that early and late phases in an auction attract more bidding activity than the middle phase. The often frenzied bidding activity at the end of an auction is referred to as *bid sniping*, and is caused by the desire of bidders to win a given auction. Besides the objective of winning, a main goal for most bidders is to win the item being auctioned at the lowest possible price.

The set of bids placed by an individual bidder during a specific auction consists of a few snapshots, taken at the times the bidder places a bid, of an underlying bidder trajectory which is a continuous function that corresponds to a specific realization of a stochastic process and reflects the bidding behavior. Our study is motivated by the goal to classify bidder trajectories based on the observed bidding activities. Classifying bidder trajectories is of interest in order to identify different bidding strategies. Bidders aim at achieving a high chance of winning and/or winning the item at a low final price. Studying how different strategies connect to these somewhat conflicting aims may help to differentiate various strategies in regard to their effectiveness to achieve these aims. As pointed out in Bapna et al. (2004), learning about bidding strategies also serves to enhance the design of online auction systems. In a recent paper by Jank and Shmueli (2008), the authors take a functional data view point and use curve clustering techniques to group the price trajectories of auctions. Their goal is to characterize heterogeneity in the price formation process of online auctions and understand sources that affect the price dynamics. Therefore, the trajectories of interest in their study are derived from the ensemble of all bids placed by all bidders who participate in a certain auction, while we are interested in the study of individual bidder-specific trajectories and the clustering of these trajectories.

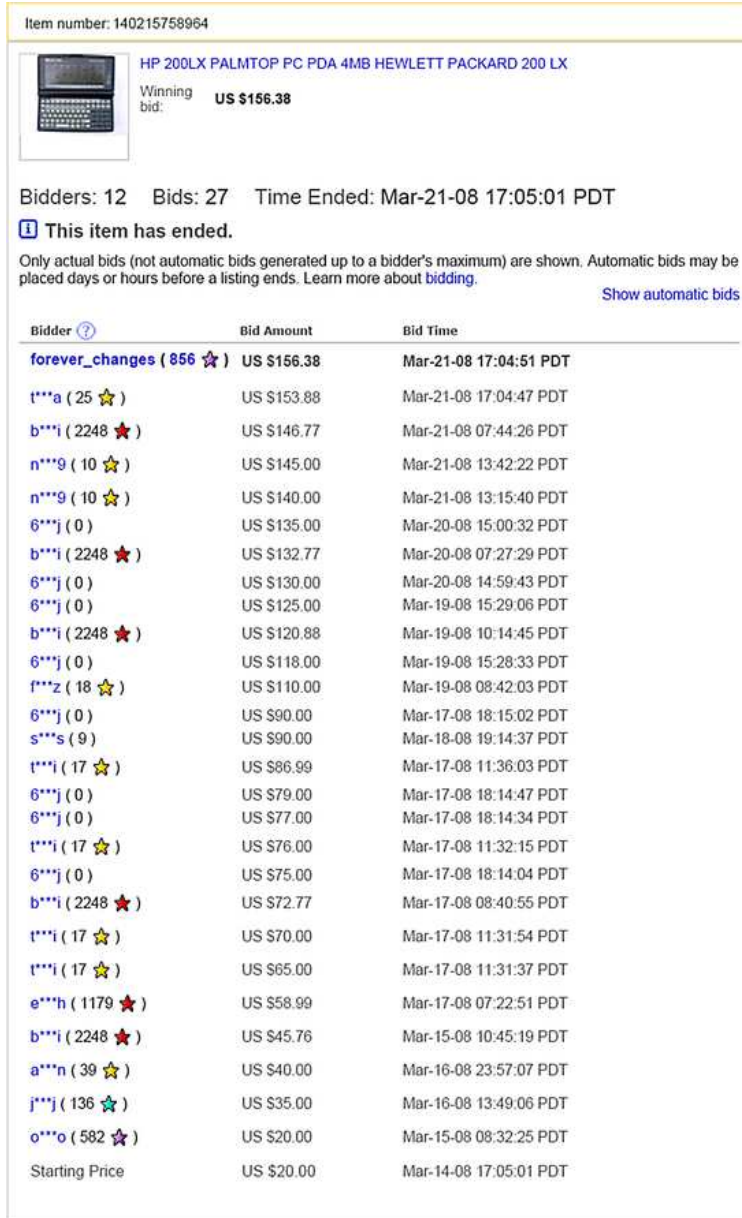

Fig. 1. *Snapshot of a seven-day eBay auction.*

A first step to differentiate between various bidding strategies is to define a distance between the various observed bidding behaviors, in order to derive a dissimilarity measure between different bidding patterns. The distance is to be based on the observed WTP bids for one bidder (in one auction), and as typical bidder trajectories are observed at only very few and irregular times, this leads to the challenge to define a distance based on sparse and irregularly timed data. Similar problems also arise in many other types of data from online environments, such as user reviews and weblog postings, where a single user might contribute a small number of entries for a specific topic/item. Assuming that an individual's bidding activity is a reflection of the realization of an underlying stochastic bid price process, this challenge motivates the development of a metric on the sample space of a stochastic process, where elements of this space are realizations of the underlying process with observations that consist of noisy measurements and are made at sparse and irregular time points. Once we have constructed a reasonable distance, we may base clustering methods on the resulting distance matrix, for example, one may apply multidimensional scaling or similar approaches. Implementing such a procedure, we find six distinct clusters of bidding patterns by analyzing the bids submitted during 158 seven-day auctions of Palm M515 Personal Digital Assistants (for more details, see Section 3). Interestingly, the chance of obtaining the auctioned item at a low price is closely associated with the bidding pattern/strategy: If the goal of the bidder is to win the auctioned item at a reasonable price, the resulting probabilities to achieve this goal show clear differences for the various bidding strategies, and one can identify better and worse strategies.

If the entire trajectory of each realized bid price process were observed, then the $L_2$ norm in the space of square integrable functions would provide a natural starting point for defining a metric. However, the $L_2$ distance is not readily calculable from the actually available noisy, sparse and irregularly sampled measurements of the bid price process. Suppose one observes a square integrable stochastic process $\{X(t) : t \in \mathcal{T}\}$ at a random number of randomly located points in $\mathcal{T}$, with measurements corrupted by additive i.i.d. random noise. The observations available from $n$ independent realizations of the process are $\{Y_{il} : 1 \le l \le n_i; 1 \le i \le n\}$ with

$$(1.1) \qquad Y_{il} = X_i(T_{il}) + \varepsilon_{il},$$

where $\{\varepsilon_{il}\}$ are i.i.d. with mean 0 and variance $\sigma^2$. Since $X$ is a square integrable stochastic process, by Mercer's theorem [cf. Ash (1972)], there exists a positive semidefinite kernel $C(\cdot, \cdot)$ such that $\text{cov}(X(s), X(t)) = C(s, t)$ and we have the following expansion of the process $X_i(t)$ in terms of the eigenfunctions of the kernel $C(\cdot, \cdot)$:

$$(1.2) \qquad X_i(t) = \mu(t) + \sum_{k=1}^{\infty} \xi_{ik} \phi_k(t),$$

where $\mu(\cdot) = \text{E}(X(\cdot))$ is the mean function; the random variables $\{\xi_{ik} : k \ge 1\}$ for each $i$ are uncorrelated with zero mean and variance $\lambda_k$; $\sum_{k=1}^{\infty} \lambda_k < \infty$, $\lambda_1 \ge \lambda_2 \ge \cdots \ge 0$ are the eigenvalues of $C(\cdot, \cdot)$; and $\phi_k(\cdot)$ are the corresponding orthonormal eigenfunctions.

In the observed data model we assume that $\{T_{il} : l = 1, \ldots, n_i; 1 \le i \le n\}$ are randomly sampled from a (possibly unknown) distribution with a density $g$ on $\mathcal{T}$. In the problems we study, $n_i$ are typically small, reflecting that the observed data consist of sparse and noisy realizations of a stochastic process. We will define a distance between two such realizations $X_i$ and $X_j$ based on the observed data as described in the next section. This approach is inspired by recent developments of functional data analysis methodology for longitudinal data, notably the work of Yao, Müller and Wang (2005) where trajectories are predicted from sparse and noisy observations, which recently was adapted to online auctions [Liu and Müller (2008)]. Approaches based on B-spline fitting with random coefficients which are suitable to fit similar data with random coefficient models have been proposed by Shi, Weiss and Taylor (1996), Rice and Wu (2001), and recently in the context of online auctions by Reithinger et al. (2008).

For an up-to-date introduction to functional data analysis, we refer to the excellent book by Ramsay and Silverman (2005). Descriptions of the rapidly evolving interface between longitudinal and functional methodology and functional models for sparse longitudinal data can be found in Rice and Wu (2001), James, Hastie and Sugar (2001), James and Sugar (2003), and the overviews provided in Rice (2004), Zhao, Marron and Wells (2004) and Müller (2005). The proposed distance is introduced in the following section. An application to the clustering of bidding patterns in eBay online auctions is the topic of Section 3, followed by concluding remarks. Proofs and auxiliary remarks can be found in an Appendix.

**2. A distance for sparse and irregular data.** We propose a distance between the random curves $X_i$ and $X_j$ based on the observed data $\mathbf{Y}_i = (Y_{i1}, \ldots, Y_{in_i})^T$ and $\mathbf{Y}_j = (Y_{i1}, \ldots, Y_{jn_j})^T$. The idea is to use the conditional expectation of the $L_2$ distance between these two curves, given the data. Our analysis is conditional on the times of the measurements $\{T_{il} : l = 1, \ldots, n_i; 1 \leq i \leq n\}$ and their numbers $\{n_i : 1 \leq i \leq n\}$.

2.1. *Definition and basic properties.* The $L_2$ distance between two curves $X_i$ and $X_j$ is defined as

$$D(i,j) = \left\{ \int_{\mathcal{T}} (X_i(t) - X_j(t))^2 \, dt \right\}^{1/2},$$

and is not calculable in our situation, as only the sparse data $\mathbf{Y}_i$ and $\mathbf{Y}_j$ are observed. Therefore, we propose to use the conditional expectation of $D^2(i,j)$, given $\mathbf{Y}_i$ and $\mathbf{Y}_j$, as the squared distance between $X_i$ and $X_j$,

$$(2.1) \qquad \tilde{D}(i,j) = \{\mathrm{E}(D^2(i,j)|\mathbf{Y}_i, \mathbf{Y}_j)\}^{1/2}, \qquad 1 \leq i, j \leq n.$$

Note that as a function of $\mathbf{Y}_i, \mathbf{Y}_j$, the $\tilde{D}(i,j)$s are random variables and have the following properties, the proof of which is given in the Appendix.

PROPOSITION 2.1. *$\tilde{D}$ satisfies the following properties:*

1. *$\tilde{D}(i,j) \geq 0$, $\tilde{D}(i,i) = 0$ and for $i \neq j$, $P(\tilde{D}(i,j) > 0) = 1$;*
2. *$\tilde{D}(i,j) = \tilde{D}(j,i)$;*
3. *For $1 \leq i, j, k \leq n$, $\tilde{D}(i,j) \leq \tilde{D}(i,k) + \tilde{D}(k,j)$.*

Therefore, $\tilde{D}$ can be viewed as a metric on the subject space consisting of random realizations $\{X_i(\cdot)\}$ of the underlying stochastic process $X(\cdot)$. Since under model (1.2), Parzeval's identity implies that the $L_2$ distance between $X_i$ and $X_j$ can be written as

$$D(i,j) = \|X_i - X_j\|_2 = \left\{ \sum_{k=1}^{\infty} (\xi_{ik} - \xi_{jk})^2 \right\}^{1/2},$$

we get

$$\tilde{D}^2(i,j) = \mathrm{E}\left( \sum_{k=1}^{\infty} (\xi_{ik} - \xi_{jk})^2 \Big| \mathbf{Y}_i, \mathbf{Y}_j \right).$$

For an integer $K \geq 1$, we then define truncated versions of $\tilde{D}$ as

$$(2.2) \qquad \begin{aligned} &\tilde{D}^{(K)}(i,j) \\ &= \left\{ \mathrm{E}\left( \sum_{k=1}^{K} (\xi_{ik} - \xi_{jk})^2 \Big| \mathbf{Y}_i, \mathbf{Y}_j \right) \right\}^{1/2} \end{aligned}$$

$$= \left\{ \sum_{k=1}^{K} \mathrm{var}(\xi_{ik}|\mathbf{Y}_i) + \mathrm{var}(\xi_{jk}|\mathbf{Y}_j) + (\mathrm{E}(\boldsymbol{\xi}_{ik}|\mathbf{Y}_i) - \mathrm{E}(\boldsymbol{\xi}_{jk}|\mathbf{Y}_i))^2 \right\}^{1/2}.$$

Note that it follows from these definitions that $\mathrm{E}(\tilde{D}^2(i,j)) = \mathrm{E}(D^2(i,j))$ and also for the truncated versions $\mathrm{E}(\tilde{D}^{(K)}(i,j)^2) = \sum_{k=1}^{K} 2\lambda_k = \mathrm{E}(D^{(K)}(i,j)^2)$, so that these conditional expectations are unbiased predictors of the corresponding squared $L_2$ distances.

2.2. *Estimation.* In the following we discuss the estimation of the truncated version of the distance $\tilde{D}^{(K)}(i,j)$ (2.2). Given an integer $K \geq 1$, let $\Lambda^{(K)} = \mathrm{diag}\{\lambda_1, \ldots, \lambda_K\}$ be the $K \times K$ diagonal matrix with diagonal elements $\{\lambda_1, \ldots, \lambda_K\}$. For $1 \leq i \leq n, 1 \leq k \leq K$, let $\boldsymbol{\mu}_i = (\mu(T_{i1}), \ldots, \mu(T_{in_i}))^T$, $\boldsymbol{\xi}_i^{(K)} = (\xi_{i1}, \ldots, \xi_{iK})^T$, $\phi_{ik} = (\phi_k(T_{i1}), \ldots, \phi_k(T_{in_i}))^T$ and $\Phi_i^{(K)} = (\phi_{i1}, \ldots, \phi_{iK})$. Define

$$(2.3) \qquad \tilde{\boldsymbol{\xi}}_i^{(K)} = \Lambda^{(K)}(\Phi_i^{(K)})^T \Sigma_{Y_i}^{-1}(\mathbf{Y}_i - \boldsymbol{\mu}_i),$$

where $\Sigma_{Y_i} = \mathrm{cov}(\mathbf{Y}_i, \mathbf{Y}_i) = (C(T_{il}, T_{il'})) + \sigma^2 I_{n_i}$. Note that $\tilde{\boldsymbol{\xi}}^{(K)}$ is the best linear unbiased predictor (BLUP) of $\boldsymbol{\xi}^{(K)}$, since $\mathrm{cov}(\boldsymbol{\xi}^{(K)}, \mathbf{Y}_i) = \Lambda^{(K)}(\Phi_i^{(K)})^T$. Moreover, if we have a finite-dimensional process, such that for some integer $K > 0$, $\lambda_k = 0$ for $k > K$ in model (1.2), then (omitting upper subscripts $K$) $\Sigma_{Y_i} = \Phi_i \Lambda (\Phi_i)^T + \sigma^2 I_{n_i}$ and $\Lambda \Phi_i^T (\Phi_i \Lambda \Phi_i^T + \sigma^2 I_{n_i})^{-1} = (\Phi_i^T \Phi_i + \sigma^2 \Lambda^{-1})^{-1} \Phi_i^T$, so that

$$\tilde{\boldsymbol{\xi}}_i = (\Phi_i^T \Phi_i + \sigma^2 \Lambda^{-1})^{-1} \Phi_i^T (\mathbf{Y}_i - \boldsymbol{\mu}_i),$$

which also is the solution of the penalized least-squares problem

$$\min_{\boldsymbol{\xi}} (\mathbf{Y}_i - \boldsymbol{\mu}_i - \Phi_i \boldsymbol{\xi})^T (\mathbf{Y}_i - \boldsymbol{\mu}_i - \Phi_i \boldsymbol{\xi}) + \sigma^2 \sum_{k=1}^{K} \xi_k^2 / \lambda_k.$$

If one assumes normality of the processes in models (1.1) and (1.2), that is, $\xi_{ik} \sim N(0, \lambda_k)$ and $\varepsilon_{il} \overset{i.i.d.}{\sim} N(0, \sigma^2)$ and independence between errors and processes, the joint distribution of $\{\mathbf{Y}_i, \boldsymbol{\xi}_i^{(K)}\}$ is multivariate normal with

$$(2.4) \qquad \begin{pmatrix} \mathbf{Y}_i \\ \boldsymbol{\xi}_i^{(K)} \end{pmatrix} \sim \mathrm{Normal}\left( \begin{pmatrix} \boldsymbol{\mu}_i \\ 0 \end{pmatrix}, \begin{pmatrix} \Sigma_{Y_i}, & \Phi_i^{(K)} \Lambda^{(K)} \\ \Lambda^{(K)}(\Phi_i^{(K)})^T, & \Lambda^{(K)} \end{pmatrix} \right).$$

Therefore, the conditional distribution of $\boldsymbol{\xi}_i^{(K)}$ given $\mathbf{Y}_i$ is normal with mean

$$(2.5) \qquad \mathrm{E}(\boldsymbol{\xi}_i^{(K)}|\mathbf{Y}_i) = \Lambda^{(K)}(\Phi_i^{(K)})^T \Sigma_{Y_i}^{-1}(\mathbf{Y}_i - \boldsymbol{\mu}_i) = \tilde{\boldsymbol{\xi}}_i^{(K)}$$

and variance

$$(2.6) \qquad \mathrm{var}(\boldsymbol{\xi}_i^{(K)}|\mathbf{Y}_i) = \Lambda^{(K)} - \Lambda^{(K)}(\Phi_i^{(K)})^T \Sigma_{Y_i}^{-1} \Phi_i^{(K)} \Lambda^{(K)}.$$

Furthermore, $\tilde{\boldsymbol{\xi}}^{(K)}$ becomes the best predictor of $\boldsymbol{\xi}^{(K)}$, and with (2.5) and (2.6),

$$
\begin{aligned}
(2.7) \quad & (\tilde{D}^{(K)}(i,j))^2 \\
&= \mathrm{tr}(\Lambda^{(K)} - \Lambda^{(K)}(\Phi_i^{(K)})^T \Sigma_{Y_i}^{-1} \Phi_i^{(K)} \Lambda^{(K)}) \\
&\quad + \mathrm{tr}(\Lambda^{(K)} - \Lambda^{(K)}(\Phi_j^{(K)})^T \Sigma_{Y_j}^{-1} \Phi_j^{(K)} \Lambda^{(K)}) \\
&\quad + \| \Lambda^{(K)}(\Phi_i^{(K)})^T \Sigma_{Y_i}^{-1}(\mathbf{Y}_i - \boldsymbol{\mu}_i) - \Lambda^{(K)}(\Phi_j^{(K)})^T \Sigma_{Y_j}^{-1}(\mathbf{Y}_j - \boldsymbol{\mu}_j) \|_2^2.
\end{aligned}
$$

Therefore, $\tilde{D}^{(K)}(i,j)$ can then be estimated by plugging in estimates for the model components, that is, for mean curve $\mu(\cdot)$, covariance kernel $C(\cdot,\cdot)$, first $K$ eigenvalues $\{\lambda_k : k = 1, \ldots, K\}$ and corresponding eigenfunctions $\{\phi_k : k = 1, \ldots, K\}$, as well as error variance $\sigma^2$. Although (2.7) is derived under the normality assumption, its expectation is always equal to the expectation of $D^{(K)}(i,j)^2$ (which is $\sum_{k=1}^{K} 2\lambda_k$), regardless of distributional assumptions.

Assuming that mean, covariance and eigenfunctions are all smooth, one can apply local linear smoothers [Fan and Gijbels (1996)], pooling observations for function and surface estimation, fitting local lines in one dimension for estimating the mean function and local planes in two dimensions for estimation of the covariance kernel [see Yao, Müller and Wang (2005) for details]. Denoting the resulting estimates of $\mu(\cdot)$, $C(\cdot,\cdot)$ by $\hat{\mu}(\cdot)$, $\hat{C}(\cdot,\cdot)$, the estimates of eigenfunctions and eigenvalues are given by the solutions $\hat{\phi}_k$ and $\hat{\lambda}_k$ of the eigen-equations

$$
\int_{\mathcal{T}} \hat{C}(s,t)\hat{\phi}_k(s)\,ds = \hat{\lambda}_k \hat{\phi}_k(t),
$$

where this system of equations is solved by discretizing the smoothed covariance [Rice and Silverman (1991)], followed by a projection on the space of symmetric and nonnegative definite covariance surfaces [Yao et al. (2003)]. The estimate $\hat{\sigma}^2$ of $\sigma^2$ is obtained by first subtracting $\hat{C}(t,t)$ from a local linear smoother of $C(t,t) + \sigma^2$, denoted by $\hat{V}(t)$, then averaging over a subset of $\mathcal{T}$ [Yao, Müller and Wang (2005)]. Further details can be found in the Appendix.

The estimate of $\tilde{D}^{(K)}(i,j)$ is then given by

$$
\begin{aligned}
(2.8) \quad \hat{D}^{(K)}(i,j) = \{ & \mathrm{tr}(\hat{\Lambda}^{(K)} - \hat{\Lambda}^{(K)}(\hat{\Phi}_i^{(K)})^T \hat{\Sigma}_{Y_i}^{-1} \hat{\Phi}_i^{(K)} \hat{\Lambda}^{(K)}) \\
& + \mathrm{tr}(\hat{\Lambda}^{(K)} - \hat{\Lambda}^{(K)}(\hat{\Phi}_j^{(K)})^T \hat{\Sigma}_{Y_j}^{-1} \hat{\Phi}_j^{(K)} \hat{\Lambda}^{(K)}) \\
& + \| \hat{\Lambda}^{(K)}(\hat{\Phi}_i^{(K)})^T \hat{\Sigma}_{Y_i}^{-1}(\mathbf{Y}_i - \hat{\boldsymbol{\mu}}_i) \\
& \quad - \hat{\Lambda}^{(K)}(\hat{\Phi}_j^{(K)})^T \hat{\Sigma}_{Y_j}^{-1}(\mathbf{Y}_j - \hat{\boldsymbol{\mu}}_j) \|_2^2 \}^{1/2},
\end{aligned}
$$

where $\widehat{\Lambda}^{(K)} = \mathrm{diag}\{\hat{\lambda}_1, \ldots, \hat{\lambda}_K\}$; $\widehat{\Phi}_i^{(K)} = (\hat{\phi}_{i1}, \ldots, \hat{\phi}_{iK})$, and the $(l, l')$ entry of $\widehat{\Sigma}_{Y_i}$ is $(\widehat{\Sigma}_{Y_i})_{l,l'} = \widehat{C}(T_{il}, T_{il'}) + \hat{\sigma}^2 \delta_{ll'}$. The computer code to calculate (2.8) based on data is given as Supplementary material [Peng and Müller (2008)]. To obtain functional principal component scores for sparse data, one can also use the matlab package PACE (http://anson.ucdavis.edu/~mueller/data/programs.html).

The following result shows the consistency of these estimates for the target distance $\tilde{D}(i, j)$, providing some assurance that the estimated distance is close to the targeted one if enough components are included and the number of observed random curves is large enough.

THEOREM 2.1. *Under the assumptions listed in Lemma A.2 in the Appendix,*

$$\lim_{K \to +\infty} \lim_{n \to +\infty} \widehat{D}^{(K)}(i, j) = \tilde{D}(i, j) \qquad in\ probability.$$

PROOF. See Appendix. □

2.3. *Distance-based scaling.* Multidimensional scaling (MDS) aims to find a projection of given original objects for which one has a distance matrix into $p$-dimensional (Euclidean) space for any $p \geq 1$, often chosen as $p = 2$ or 3 which provides best visualization. The projected points in $p$-space represent the original objects (e.g., random curves) in such a way that their distances match with the original distances or dissimilarities $\{\delta_{ij}\}$, according to some target criterion. In our setting these original distances will be the estimated conditional $L_2$ distances (2.8) between the sparsely observed random trajectories. Various techniques exist for implementing the MDS projection, including metric and nonmetric scaling.

In classical metric scaling one treats dissimilarities $\{\delta_{ij}\}$ directly as Euclidean distances and then uses the spectral decomposition of a doubly centered matrix of dissimilarities [Cox and Cox (2001)]. It is well known that there is an equivalence between principal components analysis and classical scaling when dissimilarities are truly Euclidean distances (if the subjects are points in an Euclidean space). Metric least squares scaling finds configuration points $\{x_i\}$ in a $p$-dimensional space with distances $\{d_{ij}\}$ matching $\{\delta_{ij}\}$ as closely as possible, by minimizing a loss function $S$, for example, $S = \sum_{i<j} \delta_{ij}^{-1}(d_{ij} - \delta_{ij})^2 / \sum_{i<j} \delta_{ij}$ [Sammon (1969)]. Two other popular optimality criteria for metric MDS are metric `stress` and `s-stress`, which are special cases of the criterion

$$\text{minimize} \sum_{i<j} w_{ij}[(d_{ij}^2)^r - (\delta_{ij}^2)^r]^2,$$

usually implemented with $w_{ij} = 1$. The `stress` criterion corresponds to the case $r = 1/2$ and was originally proposed by Kruskal (1964) for nonmetric

MDS, while the `s-stress` criterion corresponds to $r = 1$ and was popularized by Takane, Young and DeLeeuw (1977).

The `s-stress` criterion leads to a smooth minimization problem in contrast to `stress`. Kearsley, Tapia and Trosset (1998) applied Newton's method to find solutions using these criteria. In practice, the `stress` criterion is often normalized by the sum of squares of the dissimilarities, thus becoming scale free. Similarly, the `s-stress` criterion is normalized with the sum of the 4th powers of the dissimilarities. In the following implementation of our approach, we use metric MDS with the `s-stress` criterion for visualizing sparse and irregularly observed longitudinal data, where the MDS input distances are the estimates $\hat{D}$ (2.8) of proposed distances $\tilde{D}$ (2.7).

**3. Clustering bidders in online auctions.** Most of the eBay auctions are *second-price closed-ended* auctions with *proxy bidding*, as explained in the Introduction. In such auctions bidders submit the maximum amount that they are *willing-to-pay* (WTP) for the item. If in an ongoing auction a bidder submits a WTP amount that is higher than all previously submitted WTP amounts plus a minimum increment, then the bidder becomes the leading bidder. The winner of the auction is the leading bidder at the time the auction closes. The price that the winning bidder pays for the item is the second highest WTP plus the increment amount (second price), where the increments depend on the price level already reached. The duration of the auction is pre-determined (seven days for the auctions we consider here), and a bidder can place arbitrarily many bids at any WTP level and at any time while the auction is in progress [Shmueli and Jank (2005)]. During an auction in progress, all except the highest WTP amount are disclosed at current time, so that the "current price" observed for that auction at any given time is the second highest WTP value plus an increment. The first bid (opening bid) in an auction is set by the seller at the start of the auction and is the initial required bid amount.

We study classification of bidding behaviors using the proposed distance measure with eBay online auction data for 158 seven day auctions of Palm M515 Personal Digital Assistants (PDA) that took place between March and May, 2003. This data is publicly available at http://www.rhsmith.umd.edu/ceme/statistics/data.html. One auction which did not contain bidder ID information was removed, and the analysis reported here is based on the 157 remaining auctions. These data contain recorded bids and their respective times (where time origin is always the beginning of an auction) as well as WTP amounts for 1122 distinct bidders, 1818 different *bidder trajectories* (corresponding to the bids submitted by one bidder in one auction), for a total of 3643 submitted bids (WTP amounts). One bidder can generate a bidder trajectory in several different auctions and one auction usually consists of several bidder trajectories by several different bidders.

Among the 1818 observed bidder trajectories, 1046 consist of only one bid, and 361 of two bids. The average number of bids in a bidder trajectory is 2 with a standard deviation of 1.82. It is quite typical in eBay auctions that most bidders submit only very few bids. This results in the sparseness of observable bids corresponding to one bidder trajectory, and furthermore, the bids are irregularly placed during an ongoing auction. Individual bidder trajectories are shown in Figure 2, where the observed WTP amounts are simply connected by straight lines. The $i$th bidder trajectory consists of the series of bid times $T_{ij}$, $j = 1, \ldots, n_i$ (measured in hours, where the origin of time corresponds to the beginning of the auction), and the corresponding WTP values $Y_{ij}$ (measured in dollars). The bidder trajectory is assumed to be generated by a latent continuous random trajectory of which the observed bids are just a snapshot. Given these bidder trajectories for the $n = 1122$ bidders, we aim at an empirical classification of the bidding strategies that bidders employ.

The mean trajectory for the observed 1818 bidder trajectories is obtained by pooling all data as described in the Appendix. In a preprocessing step, using the pooled data, the residuals from the overall mean curve are calculated, and the standard deviation calculated from all such residuals is then used to remove bidder trajectories that contain outlying bids. These are defined as bids that fall outside of three standard deviations of the mean curve. We found 47 outlying bids in 17 of the bidder trajectories which were removed, leaving $n = 1801$ bidder trajectories from 1112 distinct bidders and all 157 auctions in the analysis. The total remaining number of bids is 3596. The outliers were removed to assure more robust estimation of mean curve and covariance surface. As there are only 17 bidder trajectories removed, the final clustering results are not much affected by removing the outliers. The range of bid times is from $0.18h$ to $168h$, where the time domain of each auction is $[0, 168h]$ (i.e., seven-day auctions). The data are then modeled as generated from 1801 realizations of an underlying latent stochastic bid price process, as described in Section 1. The mean trajectory (Figure 2) and covariance surface of this bid price process are estimated as described in the Appendix, with bandwidths selected by leave-one-curve-out cross validation. Even though bidder trajectories from the same auction or from the same bidder are most likely dependent to some extent, we do not expect the mild dependence to have much impact on the estimation procedures, and previous analyses such as the one reported in Bapna et al. (2004) also have used all available bidder trajectories.

We chose $K = 5$ components when estimating the conditional $L_2$ distance $\tilde{D} = (\tilde{D}(i,j))_{1 \leq i,j \leq 1801}$ [see equation (2.1)], based on the fact that these accounted for about 95% of total variation of the random trajectories. The estimated first five eigenvalues are $191.4 \times 10^3$, $30.75 \times 10^3$, $5.382 \times 10^3$, $4.183 \times 10^3$ and $2.111 \times 10^3$, explaining 78.6%, 12.6%, 2.21%, 1.72%, 0.87%

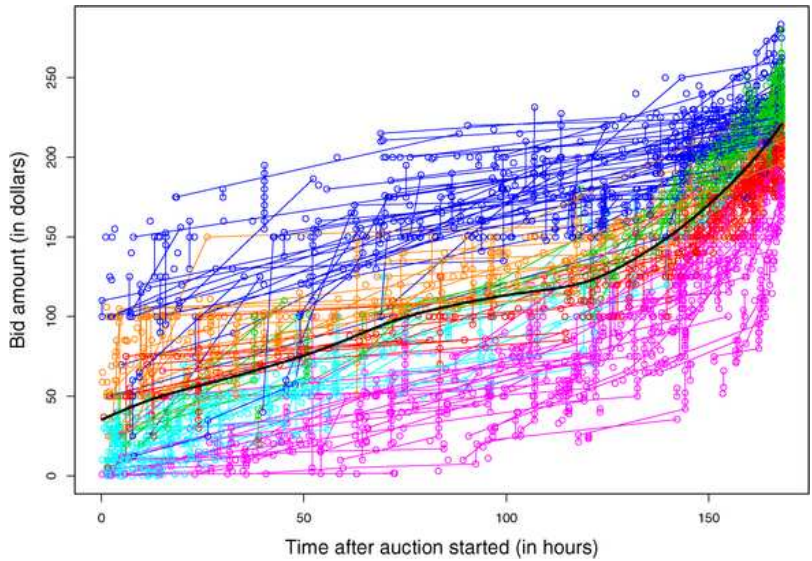

Fig. 2. *Individual bidder trajectories and mean bidder trajectory (black solid line) for* 1801 *bidder trajectories in* 157 *eBay online auctions. Individual trajectories are colored according to their cluster membership (see Figure* 3 *for the correspondence between color and cluster).*

of the sum of the first 100 eigenvalues, respectively. The estimated error variance is 697, where all amounts are in U.S. dollars. Multidimensional scaling (MDS) was then applied to the distance matrix $\tilde{D}$, projecting into a space of dimension $p = 2$, and using the matlab function `mdscale`. For the goodness of fit criterion that the MDS algorithm minimizes, we considered the criteria `sammon`, `metricstress` and `metricsstress`. Among these, only `metricsstress` converged for the data at hand. The MDS projection result is displayed in Figure 3. This figure clearly reveals that the bidder trajectories can be separated into distinct subgroups.

Applying K-means cluster analysis to the `metricsstress` MDS results, six clusters of bidder trajectories can be identified (Figure 3). We label these clusters as "L" (*Low end*), "H" (*High end*), "S" (*Slow start*), "F" (*Fast start, slow increase*), "A" (*Aggressive*) and "E" (*Early*), respectively. Each cluster consists of a number of bidder trajectories ranging between 212 and 399. One finds that bidders with bidder trajectories in clusters "L", "H" and "E" tend to place relatively fewer bids than those in clusters "S" and "A", with those in cluster "F" placing a moderate number of bids (Table 1). Further characteristics of these six clusters regarding bid times, bid amounts and chances of winning are summarized in Table 1.

Figures 4 and 5 display the line connected individual bidder trajectories and the fitted mean trajectories for the six clusters of bidder trajectories.

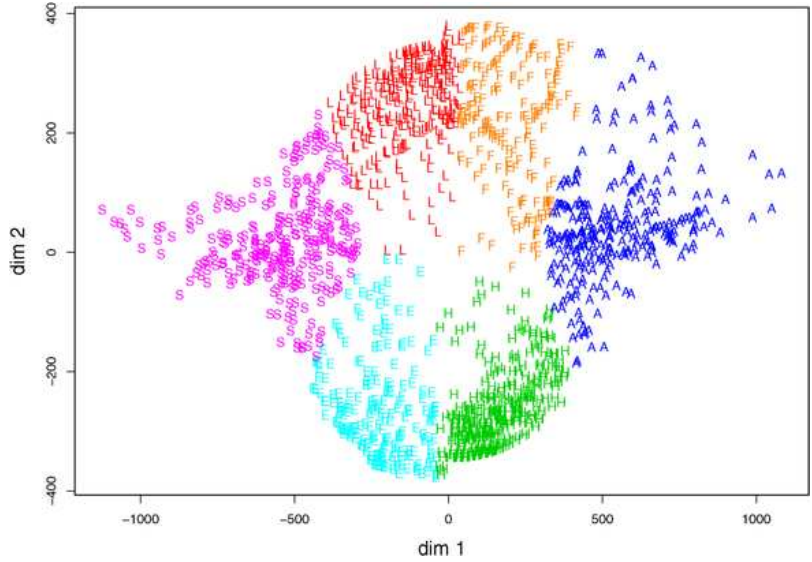

FIG. 3. *Multidimensional scaling (MDS) using the criterion* `metricsstress` *applied to the proposed conditional functional distance and projecting to two-dimensional space. Also shown is the K-means clustering result which reveals six clusters: "L" (Low end), "H" (High end), "S" (Slow start), "F" (Fast start, slow increase), "A" (Aggressive) and "E" (Early).*

TABLE 1
*Summary statistics for the six clusters*

| Group ID | Number of bidder trajectories | Average number of bids (std) | Bid times $(Q_1, m, Q_3)^*$ | Bid amounts $(Q_1, m, Q_3)^*$ | Number of winners (%) | Average amount paid at winning (std) |
|---|---|---|---|---|---|---|
| "L" | 386 | 1.80 (1.60) | (132, 158, 166) | (125, 175, 200) | 29 (7.51%) | 208 (10.8) |
| "H" | 399 | 1.76 (1.36) | (155, 165, 168) | (200, 220, 233) | 97 (24.3%) | 237 (11.2) |
| "S" | 298 | 2.51 (2.39) | (83, 124, 152) | (45, 77, 122) | 2 (0.67%) | 177 (10.4) |
| "F" | 212 | 1.97 (1.93) | (22, 58, 110) | (80, 101, 140) | 0 (0%) | NA (NA) |
| "A" | 285 | 2.19 (1.65) | (75, 123, 149) | (157, 187, 215) | 26 (9.12%) | 245 (16.5) |
| "E" | 221 | 1.85 (1.94) | (8.4, 22, 54) | (13, 35, 61) | 1 (0.45%) | 230 (NA) |
| Overall | 1801 | 2 (1.82) | (65, 136, 162) | (77, 150, 200) | 155 (8.6%) | 232 (18.3) |

$^*(Q_1, m, Q_3)$ stands for (first quartile, median, third quartile).

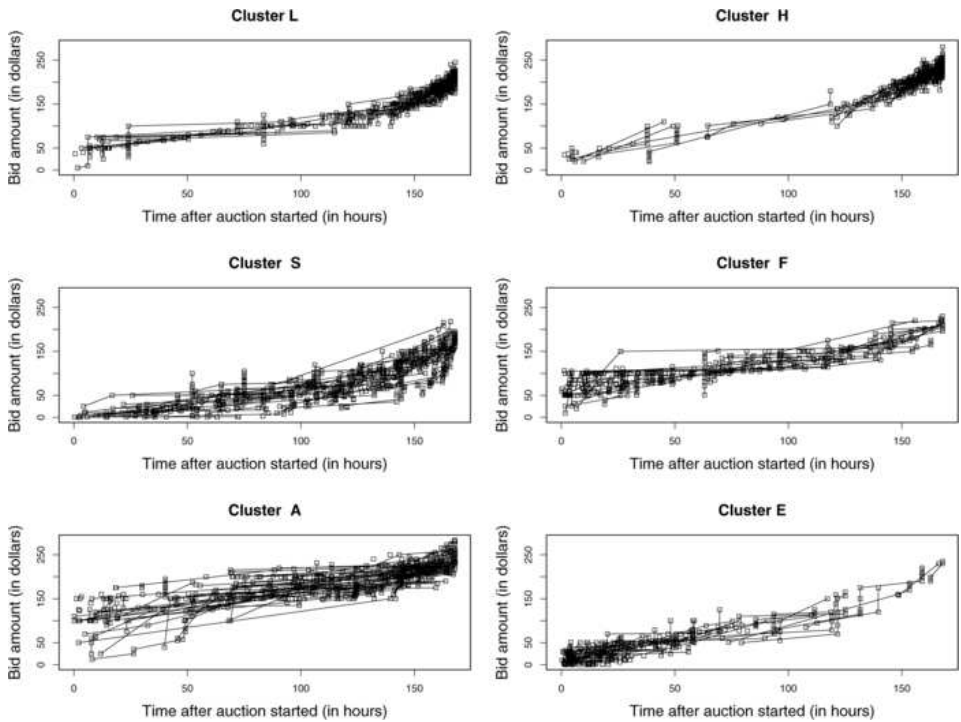

Fig. 4.  *Individual bidder trajectories for each of the six clusters.*

The histograms of the bid time and bid amount for each cluster are visualized in Figures 6 and 7. One finds that "L" and "H" bidders tend to place bids over a short period of time near the end of an auction, and participate in "bid sniping." As can be seen from Figure 7, near the end of an auction, "L" bidders place relatively low bids, so these bidders aim at a good bargain, while "H" bidders place relatively high bids, so the primary interest of these bidders seems to be to secure the item, while price plays a secondary role. In contrast to "L" and "H" bidders, "E" bidders tend to bid at the beginning of an auction, placing low bids, and then refrain from placing subsequent competitive bids. "S", "F" and "A" bidders tend to place bids throughout the entire auction. Among these, "S" bidders tend to slowly raise their bids at the beginning and then increase them faster toward the end of the auction, giving rise to a convex mean curve. "F" bidders, on the other hand, place relatively high bids at the beginning (mostly starting around 50 dollars), and from then on only cautiously raise their bids, ending their bidding at a fairly modest price level. Their mean bidder trajectory is almost linear. While "A" bidders also start with high bids at the beginning, they increase their bids more aggressively compared to the "F" bidders throughout the auction and end up at a high price level.

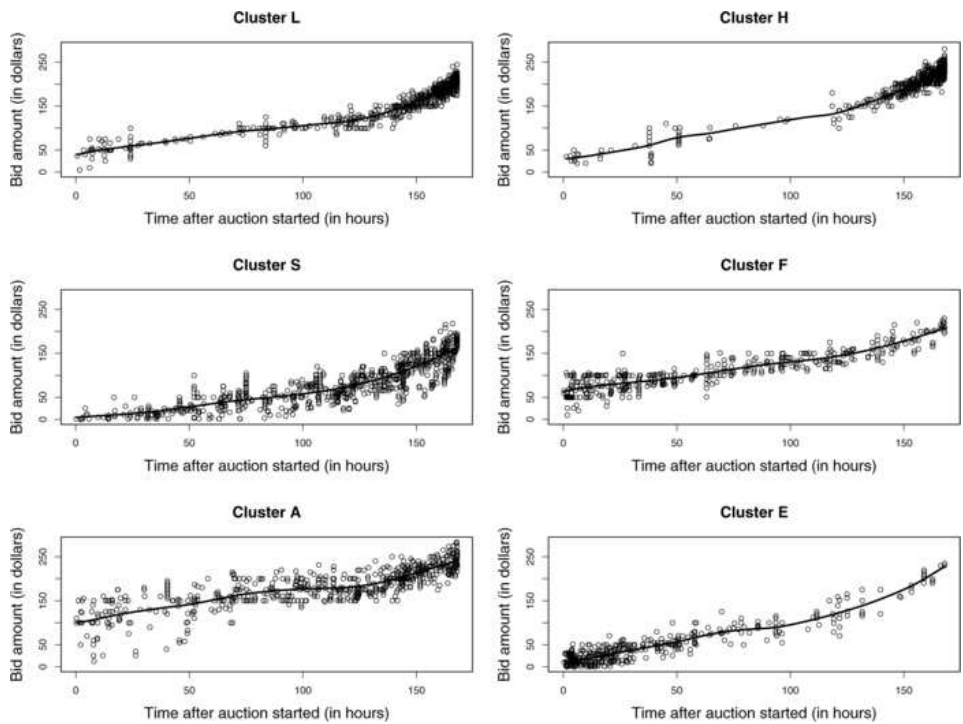

Fig. 5. *Mean bidder trajectory (black solid line) for each of the six clusters.*

It is instructive to study the winning rate for each cluster. The winning rate for a given cluster is defined as the proportion of the winning trajectories, that is, the fraction of the bidders in the cluster winning the auction. We find that "H" bidders have the highest winning rate: around 24.3%; this is not too surprising as these bidders tend to place high bids near the end of an auction. The success rates of "A" bidders and "L" bidders are lower and comparable to each other, 9.1% and 7.5%, respectively. The chance of winning the auction is very slim for the remaining three groups. Indeed, "S" and "E" groups only include 2 and 1 winners respectively, and there is no winner in the "F" group. By examining the average prices paid when winning the bid, "A" bidders pay highest with 245$, while "H" bidders pay slightly less with 237$. "L" bidders pay considerably less with 208$, while "E" bidders and "S" bidders pay 230$ and 177$, respectively. For more details, see Table 1.

Our results confirm the effectiveness of the "bid sniping" strategy in eBay online auctions. If the goal is to secure the item, then "H" emerges as the best strategy. If, however, the goal is to win the item with a low price and a reasonable chance, then "L" is the best strategy. Since both strategies involve a fewer number of bids, they are also best in terms of the efforts invested in an auction by a bidder.

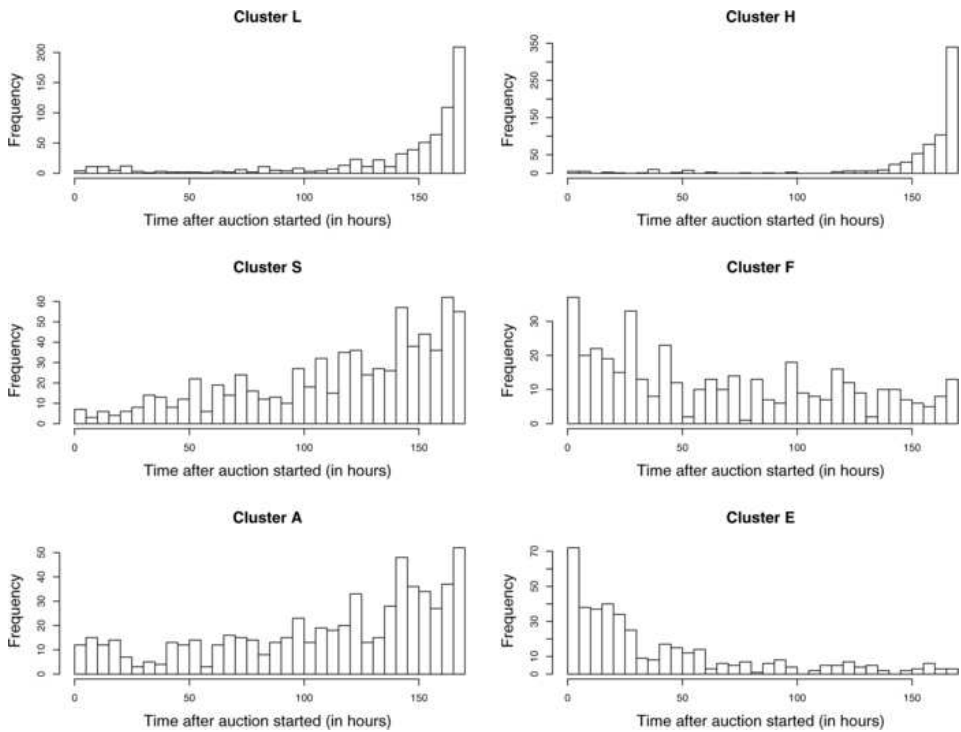

Fig. 6.    *Histogram of times when bids are placed for each of the six clusters.*

**4. Discussion.**  Bapna et al. (2004) proposed to use clustering of bidders based on three summary statistics: time of entry, time of exit and number of bids. It is well known that the beginning and ending of an online auction usually attract more bids than the middle period and these three statistics can be expected to partially reflect the timing of the bidding. Clustering based on these three statistics has been applied to multiunit Yankee auctions which use a format that is quite different from the format used in eBay auctions. These statistics have been shown to lead to sensible and insightful clustering results for this format. However, these statistics do not reflect the actual bid amounts placed during an auction, and one can expect that including the amounts adds valuable information that leads to improved clustering.

In fact, the six bidding strategies we identified are not only differentiated by the timing when the bids are submitted, but even more so by the bid amounts and especially the interaction between the two, namely, at which time when lower, when higher and when no bids at all are submitted. These interactions carry important information that characterizes bidding strategies. For example, when ignoring the actual bid amounts, one could not distinguish between "L" and "H" bidders, as they show quite similar patterns in terms of timing when bids are placed, but differ substantially in the amounts of the bids placed. The functional viewpoint that we advocate reveals more features of a particular strategy, as it allows a natural description of the interactions between the timing of bids and their amounts, for example, in the characterization of the "S", "A" and "F" clusters.

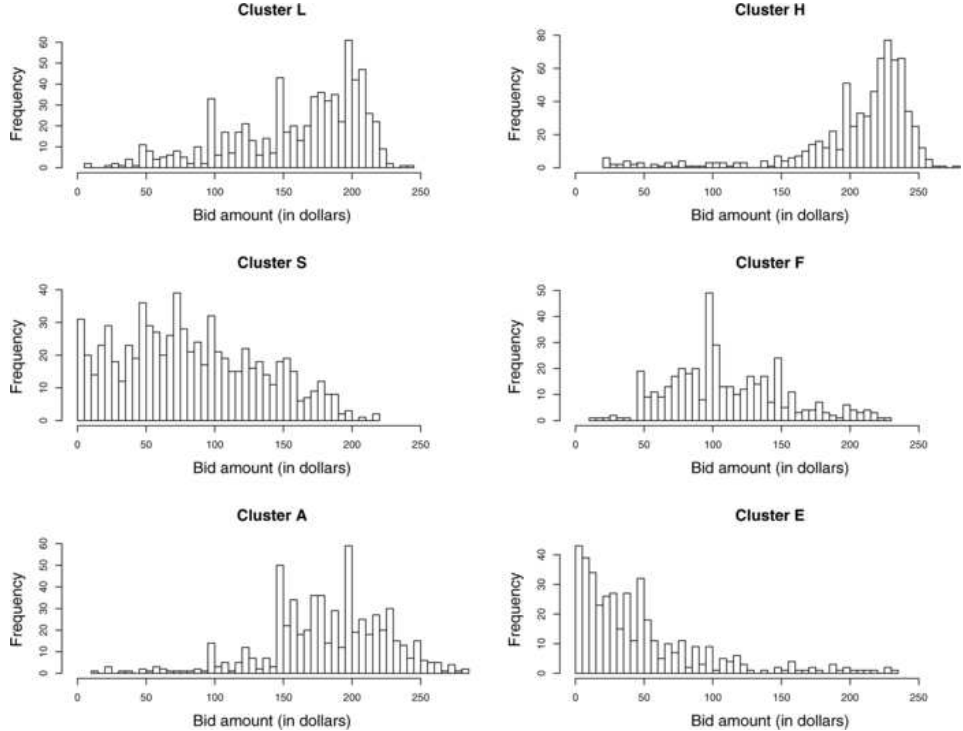

Fig. 7. *Histogram of bid (willingness-to-pay) amounts for each of the six clusters.*

The proposed functional distance is defined conditionally on observed "snapshots" of the underlying trajectories. It provides a useful tool and solves the problem to defining a distance when the data are sparse and irregular. After the distances for such data have been obtained, one can apply standard tools such as MDS and clustering methods to classify such data. We demonstrated that the estimated distances converge to target values under suitable assumptions.

Applying these methods to online auction data leads to insights on bidding strategies. For the second-price closed-ended eBay online auction data that we have considered, bidding behaviors fall into six distinct clusters, each associated with specific characteristics of the corresponding bidder trajectories. Each of these bidding strategies is associated with a distinct chance to win the item that is auctioned, and also with a distinct final price that the winner of the auction has to pay for the item. It turns out that the six bidding strategies can be clearly distinguished in terms of how well they achieve these goals. The strategy of placing bids near the end of an auction at moderate bid levels ("L" cluster) is most effective when one aims at combining a reasonable chance of winning with a relatively low price. The proposed methodology is more generally useful for all longitudinal studies where clustering of subjects is of interest and data are sparse and irregular.

## APPENDIX

PROOF OF PROPOSITION 2.1.   The first two properties are obvious from the definition of $\tilde{D}$, while the third one can be easily proved by applying the Cauchy–Schwarz inequality. Note that $D$ is a metric, thus satisfying the triangle inequality: $D(i, j) \leq D(i, k) + D(k, j)$. Therefore,

$$D^2(i, j) \leq D^2(i, k) + D^2(k, j) + 2D(i, k)D(k, j).$$

Since $\mathrm{E}(D^2(i, j)|\mathbf{Y}_i, \mathbf{Y}_j, \mathbf{Y}_k) = \mathrm{E}(D^2(i, j)|\mathbf{Y}_i, \mathbf{Y}_j) = \tilde{D}^2(i, j)$, we then have

$$\tilde{D}^2(i, j) \leq \tilde{D}^2(i, k) + \tilde{D}^2(k, j) + 2\mathrm{E}(D(i, k)D(k, j)|\mathbf{Y}_i, \mathbf{Y}_j, \mathbf{Y}_k)$$

and by the Cauchy–Schwarz inequality,

$$\begin{aligned}
\mathrm{E}(D(i, k)&D(k, j)|\mathbf{Y}_i, \mathbf{Y}_j, \mathbf{Y}_k) \\
&\leq \{\mathrm{E}(D^2(i, k)|\mathbf{Y}_i, \mathbf{Y}_j, \mathbf{Y}_k)\}^{1/2}\{\mathrm{E}(D^2(k, j)|\mathbf{Y}_i, \mathbf{Y}_j, \mathbf{Y}_k)\}^{1/2} \\
&= \tilde{D}(i, k)\tilde{D}(j, k).
\end{aligned}$$

This concludes the proof.   □

**Model fitting.**   Following Yao, Müller and Wang (2005), we fit the model by local linear smoothers based on pooled data. For the mean curve $\mu(\cdot)$, applying weighted least squares, one obtains

$$(A.1) \quad (\hat{\beta}_0, \hat{\beta}_1) = \arg\min_{\beta_0, \beta_1} \sum_{i=1}^n \sum_{j=1}^{n_i} K_1\left(\frac{T_{ij} - t}{h_\mu}\right)\{Y_{ij} - \beta_0 - \beta_1(t - T_{ij})\}^2,$$

where $K_1$ is a one-dimensional kernel, for example, the univariate Epanechnikov kernel $K_1(x) = \frac{3}{4}(1 - x^2)I_{[-1,1]}(x)$, and $h_\mu$ is the bandwidth which can be selected by leave-one-curve-out cross validation. The resulting estimate is $\hat{\mu}(t) = \hat{\beta}_0(t)$. The covariance kernel $C(\cdot, \cdot)$ is fitted by two-dimensional weighted least squares,

$$
\begin{aligned}
(A.2) \quad & (\hat{\beta}_0, \hat{\beta}_1, \hat{\beta}_2) \\
&= \arg\min_{\beta_0, \beta_1, \beta_2} \sum_{i=1}^{n} \sum_{1 \le j \neq l \le n_i} K_2\left(\frac{T_{ij} - s}{h_G}, \frac{T_{il} - t}{h_G}\right) \\
&\qquad\qquad\qquad \times \{G_i(T_{ij}, T_{il}) - f(\beta, s, t, T_{ij}, T_{il})\}^2,
\end{aligned}
$$

where $G_i(T_{ij}, T_{il}) = (Y_{ij} - \hat{\mu}(T_{ij}))(Y_{il} - \hat{\mu}(T_{il}))$, $f(\beta, s, t, T_{ij}, T_{il}) = \beta_0 + \beta_1(s - T_{ij}) + \beta_2(t - T_{il})$, $K_2$ is a two-dimensional kernel, for example, the bivariate Epanechnikov kernel $K_2(x, y) = \frac{9}{16}(1 - x^2)(1 - y^2)I_{[-1,1]}(x)I_{[-1,1]}(y)$ and $h_G$ is a bandwidth which can be selected by leave-one-curve-out cross validation. Then $\hat{C}(s, t) = \hat{\beta}_0(s, t)$, and we note that since $E(G_i(T_{ij}, T_{il})) \approx C(T_{ij}, T_{il}) + \sigma^2 \delta_{jl}$, in (A.2) one should only use the off diagonal entries of the empirical covariance, that is, $G_i(T_{ij}, T_{il})$, $j \neq l$. The function $V(t) = C(t, t) + \sigma^2$ is fitted by

$$
\begin{aligned}
(A.3) \quad & (\hat{\beta}_0, \hat{\beta}_1) = \arg\min_{\beta_0, \beta_1} \sum_{i=1}^{n} \sum_{j=1}^{n_i} K_1\left(\frac{T_{ij} - t}{h_V}\right) \\
&\qquad\qquad\qquad \times \{G_i(T_{ij}, T_{ij}) - \beta_0 - \beta_1(t - T_{ij})\}^2,
\end{aligned}
$$

where $K_1$ is the one-dimensional kernel, $h_V$ a bandwidth, and $\hat{V}(t) = \hat{\beta}_0(t)$. Then the estimate of the error variance $\sigma^2$ is given by

$$
(A.4) \qquad \hat{\sigma}^2 = \frac{2}{\mathcal{T}} \int_{\mathcal{T}_1} (\hat{V}(t) - \hat{C}(t, t)) \, dt,
$$

where $\mathcal{T}_1$ is the middle half of the interval $\mathcal{T}$.

We next state a series of auxiliary lemmas. The first lemma summarizes asymptotic results from Yao, Müller and Wang (2005). The set of assumptions (A1.1)–(A4) and (B1.1)–(B2.2b) is given in Yao, Müller and Wang (2005) and will not be repeated here.

LEMMA A.1 [Theorem 1, Corollary 1 and Theorem 2 of Yao, Müller and Wang (2005)]. *Under* (A1.1)–(A4) *and* (B1.1)–(B2.2b) *with* $\nu = 0$, $\ell = 2$ *in* (B2.2a) *and* $\nu = (0, 0)$, $\ell = 2$ *in* (B2.2b),

$$
\sup_{t \in \mathcal{T}} |\hat{\mu}(t) - \mu(t)| = O_p\left(\frac{1}{\sqrt{n} h_\mu}\right), \qquad \sup_{t, s \in \mathcal{T}} |\hat{C}(s, t) - C(s, t)| = O_p\left(\frac{1}{\sqrt{n} h_G^2}\right),
$$

$$
|\hat{\sigma}^2 - \sigma^2| = O_p\left(\frac{1}{\sqrt{n}}\left(\frac{1}{h_G^2} + \frac{1}{h_V}\right)\right), \qquad |\hat{\lambda}_k - \lambda_k| = O_p\left(\frac{1}{\sqrt{n} h_G^2}\right),
$$

$$
\|\hat{\phi}_k - \phi_k\|_H = O_P\left(\frac{1}{\sqrt{n} h_G^2}\right), \qquad \sup_{t \in \mathcal{T}} |\hat{\phi}_k(t) - \phi_k(t)| = O_p\left(\frac{1}{\sqrt{h} h_G^2}\right).
$$

LEMMA A.2.  *Under the normality assumption and the set of assumptions as in Lemma A.1,*

$$\lim_{n \to +\infty} \widehat{D}^{(K)}(i,j) = \tilde{D}^{(K)}(i,j) \qquad in\ probability.$$

PROOF.   Recall that $\widehat{D}^{(K)}(i,j)$ is given by (2.8), which under the normality assumption equals $\tilde{D}^{(K)}(i,j)$ with unknown model components replaced by their estimates. The result follows from Lemma A.1 and Slutsky's theorem.  □

LEMMA A.3.

$$\lim_{K \to \infty} \tilde{D}^{(K)}(i,j) = \tilde{D}(i,j) \qquad in\ probability.$$

PROOF.   By definition,

$$\tilde{D}^2(i,j) - \tilde{D}^{(K)}(i,j)^2 = \mathrm{E}\left( \sum_{k=K+1}^{\infty} (\xi_{ik} - \xi_{jk})^2 \Big| \mathbf{Y}_i, \mathbf{Y}_j \right) \geq 0.$$

Thus,

$$\mathrm{E}(\tilde{D}^2(i,j) - \tilde{D}^{(K)}(i,j)^2) = \mathrm{E}\left( \sum_{k=K+1}^{\infty} (\xi_{ik} - \xi_{jk})^2 \right).$$

Note that $\xi_{ik}$ and $\xi_{jk}$ are independent with mean zero and variance $\lambda_k$, and therefore,

$$\mathrm{E}\left( \sum_{k=K+1}^{\infty} (\xi_{ik} - \xi_{jk})^2 \right) = 2 \sum_{k=K+1}^{\infty} \lambda_k.$$

Since $\sum_{k=1}^{\infty} \lambda_k < \infty$, then

$$\lim_{K \to \infty} \sum_{k=K+1}^{\infty} \lambda_k = 0.$$

Therefore, by Markov's inequality and the fact that $\tilde{D}^2(i,j) - \tilde{D}^{(K)}(i,j)^2 \geq 0$, for any $\epsilon > 0$,

$$P(|\tilde{D}^2(i,j) - \tilde{D}^{(K)}(i,j)^2| > \epsilon) \leq \mathrm{E}(\tilde{D}^2(i,j) - \tilde{D}^{(K)}(i,j)^2)/\epsilon$$

$$= 2 \sum_{k=K+1}^{\infty} \lambda_k/\epsilon \longrightarrow 0.$$

The result follows from Slutsky's theorem.  □

PROOF OF THEOREM 2.1.   Note that

$$|\widehat{D}^{(K)}(i,j) - \tilde{D}(i,j)| \leq |\widehat{D}^{(K)}(i,j) - \tilde{D}^{(K)}(i,j)| + |\tilde{D}^{(K)}(i,j) - \tilde{D}(i,j)|.$$

By Lemma A.3, for any $\epsilon > 0$ and any $\delta > 0$, there exists $K_0$ such that, for any $K \geq K_0$,

$$P(|\tilde{D}^{(K)}(i,j) - \tilde{D}(i,j)| \geq \epsilon/2) \leq \delta/2.$$

By Lemma A.2, for each $K > 0$, there exists $n_0(K) > 0$ such that, for any $n \geq n_0(K)$,

$$P(|\widehat{D}^{(K)}(i,j) - \tilde{D}^{(K)}(i,j)| \geq \epsilon/2) \leq \delta/2.$$

Therefore, for $K \geq K_0$, $n \geq n_0(K)$, $P(|\widehat{D}^{(K)}(i,j) - \tilde{D}(i,j)| \geq \epsilon) \leq \delta$, which concludes the proof. □

**Acknowledgments.** We are grateful to Wolfgang Jank for sharing the eBay auction data, and to one reviewer whose comments led to many improvements of the paper.

## SUPPLEMENTARY MATERIAL

**Supplement A: eBay codes** (DOI: 10.1214/08-AOAS172SUPPA; .txt).

**Supplement B: R functions used for FPCA and conditional distance analysis** (DOI: 10.1214/08-AOAS172SUPPB; .txt). These functions are used in eBay_codes.txt.

Department of Statistics
University of California
Davis, California 95616
USA
E-mail: jie@wald.ucdavis.edu
        mueller@wald.ucdavis.edu